\documentclass [preprint,showpacs,nofootinbib,superscriptaddress]{revtex4}
\usepackage{verbatim}
\usepackage{slashed}
\usepackage{epsfig}
\usepackage{amsmath}
\usepackage{mathrsfs}
\usepackage{bm}

\begin{document}

\preprint{DESY~13--198 \hspace{12.6cm} ISSN 0418-9833}
\preprint{November 2013\hspace{15.2cm}}


\title{Search for $C=+$ charmonium states in
$e^+e^-\to \gamma+~X$ at BEPCII/BESIII}



\author{Kuang-Ta Chao}
\affiliation{Department of Physics and State Key Laboratory of
Nuclear Physics and Technology, Peking University, Beijing 100871,
China}

\author{Zhi-Guo He}
\affiliation{II. Institut f\"{u}r Theoretische Physik, Universit\"{a}t Hamburg,
Luruper Chaussee 149, 22761 Hamburg, Germany}

\author{Dan Li}
\affiliation{School of Science, Zhejiang University of Science and Technology,
Hangzhou, 310023, China}

\author{Ce Meng}
\affiliation{Department of
Physics and State Key Laboratory of Nuclear Physics and Technology,
Peking University, Beijing 100871, China}


\begin{abstract}
We extend our original study in Ref.~\cite{LiDan2009} on the production of
$C=+$ charmonium states $X=\eta_c(1S/2S)$ and $\chi_{cJ}(1P/2P)$ in $e^+e^-
\to\gamma~+~X$ at B factories to the BEPCII/BESIII energy region with
$\sqrt{s}=4.0\mbox{-}5.0$ GeV. In the framework of nonrelativistic QCD factorization,
the cross sections are estimated to be as large as $0.1\mbox{-}0.9$
pb. The results could be used to search for the missing $2P$ charmonium states or
to estimate the continuum backgrounds in the resonance region.
\end{abstract}


\maketitle


\section{Introduction}

In the last ten years there have been a number of exciting discoveries of new hidden
charm states, i.e., the so-called $XYZ$ mesons (see Ref.~\cite{QWG:2011review}
for a comprehensive review). Among these states, some of the $C=+$ states around
3.9 GeV may be the candidates for the missing $2P$ charmonia. E.g., the $Z(3930)$
should be $\chi_{c2}(2P)=\chi_{c2}'$ according to its production rate and
quantum numbers measured by the Belle Collaboration~\cite{Belle06:Z3930}, and
the $X(3872)$ could be a mixed state of the $\chi_{c1}(2P)=\chi_{c1}'$
and the $D^0\bar{D}^{*0}+c.c.$ continuum as suggested in Ref.\cite{Meng-BtoX3872}.

Four years ago, we proposed to search for the missing $2P$ charmonium
states in the process $e^+e^-\to\gamma\chi_{cJ}(2P)$ and to further identify
the nature of the relevant $XYZ$ states~\cite{LiDan2009}. The cross section of
the process is calculated within the framework of NRQCD factorization~\cite{Bodwin:1994jh} at the next-to-leading order (NLO) in $\alpha_s$,
and the results are consistent with a similar but independent calculation in
Ref.~\cite{SC}. Phenomenologically, we evaluated the cross sections at the center
of mass (c.o.m.) energy at $B$-factories, i.e., $\sqrt{s}=10.6$ GeV, and found
that they are no more than several tens fb~\cite{LiDan2009}. Needless to say,
the same processes can also be used to search for these states at BEPCII/BESIII
with the c.o.m energy $\sqrt{s}$ above 4 GeV. Since the cross sections
roughly scale as $1/s^2$ with the energy, they should be significantly enhanced at BEPCII/BESIII
as compared with those at the $B$-factories.

However, there are many vector resonances lying in the energy region above 4 GeV,
such as $\psi(4040,4160,4415),\ Y(4260,4350,4660)$ etc., which can decay to the
$2P$ charmonium states through the electric dipole (E1) transitions between
charmonia~\cite{Li2012-E1} or some other exotic mechanisms. To clarify the situation,
one need to separate the resonance contributions from the non-resonance ones.
From this point of view, it is also necessary to reevaluate the cross section of $e^+e^-\to\gamma
\chi_{cJ}(2P)$ at the BEPCII/BESIII energy region to estimate the magnitude of the
non-resonance contribution.

On the other hand, by compared with the theoretical results, the measurements of
the cross sections $e^+e^-\to\gamma X$, where $X$ denotes any $C=+$ charmonia, would
also be used to testify the production mechanism and the universality of the NRQCD
long-distant matrix elements (LDMEs), especially when the resonance contributions
are not important. This could be the case for the production of $\eta_c^{(\prime)}$
since in the nonrelativistic case, the amplitudes of the magnetic-dipole (M1)
transitions between $\eta_c^{(\prime)}$ and higher vector chamonia are strongly
suppressed. Therefore, we will also evaluated the cross section for the production
of $\eta_c^{(\prime)}$ at the BEPCII/BESIII energy region.

We organize our paper as follows. In Section II, we will briefly review the framework
of the calculations. The numerical results and the phenomenology discussions will be presented
in section III. The last section is a short summary.

\section{Framework of Calculation}

Based on the NRQCD factorization formula \cite{Bodwin:1994jh}, the amplitude for
$e^{+}e^{-}\to \gamma+ X(^{2S+1}L_{J})$ can be expressed as
\begin{eqnarray}
&&\mathcal{M}(e^{+}e^{-}\to\gamma+X)=\sum_{S,L}\sum_{s_1,s_2}\sum_{i,j}
\int
\frac{{d}^3\mathbf{q}}{(2\pi)^{3}2q^{0}}\delta(q^{0}-\frac{\mathbf{q}^2}{2m_c})\psi_{LL_z}
(\mathbf{q})\langle s_1,\hspace{-0.1cm}s_2|SS_z\rangle\nonumber\\
&&\langle LL_z,SS_z|JJ_z\rangle\langle i,j|1\rangle
\mathcal{A}(e^{+}e^{-}\rightarrow\gamma+
c_{s_1}^{i}(\frac{P}{2}+q)+\overline{c}_{s_2}^{j}(\frac{P}{2}-q))
\label{amplitude}\end{eqnarray}
where $P$ is the momentum of $X$ state, $2q$ is the relative momentum between $c$
and $\overline{c}$, $\langle LL_z;SS_z|JJ_z\rangle$, $\langle s_1;s_2|SS_z\rangle$
and $\langle i,j|1\rangle=\delta_{i,j}/\sqrt{N_c}$ are the spin-SU(2), angular momentum
C-G coefficients and color-SU(3) C-G coefficient for $c\bar{c}$ pair projecting onto
appropriate bound states, respectively, and $\mathcal{A}$ is the standard Feynman
amplitude denoting $e^{+}e^{-}\rightarrow\gamma+ c_{s_1}^{i}(\frac{P}{2}+q)+
\overline{c}_{s_2}^{j}(\frac{P}{2}-q)$.

At leading order (LO), the perturbative part includes only pure QED contribution. The
cross sections can be computed straightforwardly by implementing the formulas described
in Ref.\cite{LiDan2009}. For the convenience of discussion, we list the analytical
results here, which are
\begin{subequations}\label{CS-QED}
\begin{eqnarray}
\sigma(e^+e^-\rightarrow\gamma+\eta_c)=\frac{3\alpha^3e_c^4|R_S(0)|^2(1-r)}{s^2m_c}\int
d\Omega(1+\cos^2(\theta))
\end{eqnarray}
\begin{eqnarray}
\sigma(e^+e^-\rightarrow\gamma+\chi_{c0})=\frac{3\alpha^3e_c^4|R^{\prime}_P(0)|^2(1-3r)^2}{s^2m_c^3(1-r)}\int
d\Omega(1+\cos^2(\theta))
\end{eqnarray}
\begin{eqnarray}
\sigma(e^+e^-\rightarrow\gamma+\chi_{c1})=\frac{18\alpha^3e_c^4|R^{\prime}_P(0)|^2}{s^2m_c^3(1-r)}\int
d\Omega(1+2r+(1-2r)\cos^2(\theta))
\end{eqnarray}
\begin{eqnarray}
\sigma(e^+e^-\rightarrow\gamma+\chi_{c2})=\frac{6\alpha^3e_c^4|R^{\prime}_P(0)|^2}{s^2m_c^3(1-r)}\int
d\Omega(1+6r+6r^2+(1-6r+6r^2)\cos^2(\theta))
\end{eqnarray}
\end{subequations}
where $r=4m_c^2/s$, $\theta$ is the angle between $\gamma$ and the initial $e^{+}e^{-}$
beam axis.

At QCD one-loop level, only the virtual corrections are involved. We adopt the on-shell
renormalization scheme to remove the ultraviolet divergences, in which the renormalization
constants are chosen to be
\begin{eqnarray}
&&\delta Z_2^{\rm
OS}=-\frac{1}{\varepsilon_{UV}}+\gamma_E-4-\frac{2}{\varepsilon_{IR}}-
\log(\frac{4\pi\mu^2}{m^2}),\nonumber\\
&&\delta Z_1^{\rm OS}=\delta Z_2^{\rm OS}.
\end{eqnarray}
Note that we omit the coefficient in front of the self-energy renormalization constant and
part of the infrared divergence term in $\delta Z_2^{OS}$. The cancelation of the infrared
divergences is checked both numerically and analytically, and the Coulomb singularities are
absorbed into the long-distance matrix elements, i.e. the wave functions in Eq.(\ref{CS-QED}),
through matching the results between full QCD and NRQCD calculations. More details of our
computation can be found in Ref.\cite{LiDan2009}.

Before presenting the numerical results, we would like to address some of the potential problems of the
NRQCD factorization approach. The Born cross sections in Eq.(\ref{CS-QED}) show that in the case of P-wave
production, there exists the $\frac{1}{(1-r)}$ singularity, while in the case of S-wave production it
disappears. If $m_c$ was set to be $M_{X}/2$, the cross sections for $\chi_{cJ}$ production
would be divergent near the threshold region, where $r\to 1$. One can find that the appearance of
the singularity near the threshold is due to the fact that the recoil photon is soft. It can
be easily derived that in the soft limit, the interactions between the photon and charm and
anti-charm quark are proportional to $a_1=\frac{(P+q)^{\alpha}}{P\cdot k}$ and $a_2=-\frac{(P-q)^{\alpha}}{P\cdot k}$,
respectively, where $k$ is the momentum of the soft photon. In the S-wave case the total contribution
of $a_1$ and $a_2$ terms is zero, while in the P-wave case it is non-zero. This is similar to
the un-canceled infrared divergences in the color singlet contributions to the P-wave decay \cite{Bodwin:1994jh}.
This indicates that the NRQCD factorization approach will be broken down when $r$ is close to 1.

In the NRQCD factorization formula, only $m_c$
rather than $M_x$ enters into the short-distance coefficients, and the value of $m_c$ is widely
chosen to be the current quark mass, which is in the range of $1.2 \sim 1.6~\mathrm{GeV}$. The mass
difference between $M_X$ and $2m_c$ is attributed to the non-perturbative effects. In this work,
we are concentrating on the $4\sim 5~\mathrm{GeV}$ energy region, thus the minimum value
of $1-r$ is about 0.36, which can be treated as being far from zero. Therefore, our results of the
short-distance parts are safe and the factorization should work well. On the other hand, since the masses of the X,Y,Z states are close
to $4~\mathrm{GeV}$, we make up some factors to remedy the phase space integrals as a compensation
for the calculations in the non-relativistic limit. The factor for $\eta_{c}(mS)$ production is
$\frac{(1-M^2_{\eta_c(mS)}/s)^3}{(1-4m^2_c/s)^3}$ since the $\gamma^{\ast}\to \gamma+\eta_c (mS)$
is a P-wave process, and the factor for $\chi_{cJ}(nP)$ production is  $\frac{1-M^2_{\chi_{cJ}(nP)}/s}
{1-4m^2_c/s}$ since $\gamma^{\ast}\to \gamma+\chi_{cJ} (nP)$ is predominantly an S-wave process.

\section{Numerical results and Discussions}

Now we proceed to present the numerical results. For simplicity, we refer to the Born cross
section as QED contribution and the one-loop correction as the QCD contribution. We choose
$m_c$=1.5 GeV and $\alpha_s(2m_c)$=0.26, and the values of wave functions at the origin are taken from potential
model calculations (see the results of the $B-T$-type potential in Ref.\cite{23}), which are listed
in Table I.

\begin{table}
\begin{center}
\caption{Numerical values of the radial wave functions at the origin
${|R_{nl}^{(l)}(0)|}^2$ for chamonium calculated
with the QCD (B-T) potential in Ref.\cite{23}.}
\begin{tabular}{cccccc}
\hline\hline States & $1S$ & $2S$ & $3S$ & $1P$ & $2P$ \\
\hline ${|R_{nl}^{(l)}(0)|}^2$ & 0.81 $\mbox{GeV}^{3}$ & 0.529 $\mbox{GeV}^{3}$ & 0.455 $\mbox{GeV}^{3}$
& 0.075 $\mbox{GeV}^{5}$ & 0.102 $\mbox{GeV}^{5}$ \\ \hline\hline
\end{tabular}
\label{table1}
\end{center}
\vspace{-0.5cm}
\end{table}

\subsection{$e^{+}e^{-}\to \gamma+\eta_c(mS)$}

The $\eta_{c}(1S)$ and $\eta_{c}(2S)$ have already been found for a long time. For the $\eta_{c}
(3S)$, the $X(3940)$ \cite{Abe:2007jna} is one of the candidates. If we do not take into account
the modification factor $\frac{(1-M^2_{\eta_c(mS)}/s)^3}{(1-4m^2_c/s)^3}$, the QED contribution
to the cross section of $e^{+}e^{-}\to \gamma+\eta_c$ is $1.37 - 0.83~\mathrm{pb}$ for $4.04<
\sqrt{s}<5~\mathrm{GeV}$. The QCD contribution is negative, and is about $-30\%$ of the QED ones.
The results of $\eta_c(2S)$ and $\eta_c(3S)$ can be easily obtained by replacing the wave function
of 1S with those of 2S and 3S, respectively. Thus, up to the $\alpha^3\alpha_s$ order, the cross
sections are
\begin{eqnarray}
 \sigma(e^{+}e^{-}\to \gamma+\eta_c(mS))=\Bigg{\{} \begin{array}{lll}
0.87\sim 0.54 \mathrm{pb}, \quad m=1\\
0.57\sim 0.35 \mathrm{pb}, \quad m=2\\
0.49\sim 0.30 \mathrm{pb}, \quad m=3
\end{array}
\end{eqnarray}

Setting $M_{\eta_c}=2.907~\mathrm{GeV}$, $M_{\eta_c(2S)}=3.549~ \mathrm{GeV}$ from PDG values~\cite{PDG2012}, and assuming
X(3940) is the $\eta_c(3S)$ state~\cite{Li09:ScreenedPotential}, the factor $\frac{(1-M^2_{\eta_c(mS)}/s)^3} {(1-4m^2_c/s)^3}$ is
almost 1 for $\eta_{c}$ and varies from $0.12(3.2\times10^{-4})$ to $0.47 (0.21)$ for
$m=2(3)$ in the energy range of $4.04<\sqrt{s}<5~\mathrm{GeV}$. The modification has such big effects on
$\eta_{c}(2S)$ and $\eta_{c}(3S)$ production that we treat them as the largest uncertainty sources
in our calculations and take the results before/after modification as the upper/lower bounds of our
predictions. After the modification,
the cross section of $\eta_{c}(3S)$ is very small, which may not be used to search for the $\eta_{c}(3S)$
state. However, the cross sections of $\eta_{c}(1S)$ and $\eta_{c}(2S)$ are large enough to be
measured. It should be interesting to study the $\eta_{c}(1S/2S)$ production mechanism
in the continuum energy region at BESIII. Moreover, another interesting mechanism for $\eta_{c}(mS)$ production
at BESIII, which is through the direct photon collision, was studied in Ref.~\cite{Sang:2012cp}. We
find that the cross sections of $e^{+}e^{-}\to \gamma+\eta_c(mS)$ are larger than those of  $e^{+}e^{-}\to
e^{+}e^{-}+\eta_{c}(mS)$ at least by a factor of 5 in the energy region of
$4<\sqrt{s}<5~\mathrm{GeV}$.

\subsection{$e^{+}e^{-}\to \gamma+\chi_{c0}(nP)$}

The QED contribution to $\gamma+\chi_{c0}$ production is about $120~ \mathrm {fb}$ at $\sqrt{s}=4.04~
\mathrm{GeV}$. However, the corresponding QCD contribution is about $-119~ \mathrm{fb}$, which almost cancel
the QED contribution entirely. Furthermore, we find that when $\sqrt{s}$ becomes larger the total contribution
becomes even negative. Therefore, due to the large theoretical uncertainty we will not do any phenomenological discussion for $\gamma+\chi_{c0}$ production here.

\subsection{$e^{+}e^{-}\to \gamma+\chi_{c1}(nP)$}

The QED contribution to $\gamma+\chi_{c1}(1P)$ production changes from $2.60~ \mathrm{pb} $ to
$0.68~ \mathrm{pb}$ when $\sqrt{s}$ varies from $4.04$ to $5.0~\mathrm{GeV}$, and it becomes
$50\%$ smaller after including the QCD contribution. Furthermore, if we use the modified phase
space factor with $M_{\chi_{c1}}=3.511~\mathrm{GeV}$, we obtain the QCD+QED result
\begin{equation}
\sigma(e^{+}e^{-}\to \gamma+\chi_{c1})=0.70 - 0.25\; (\mathrm{pb})
\quad \mathrm{for}\;\; 4.04<\sqrt{s}<5\; \mathrm{GeV.}
\end{equation}
Unlike the $\eta_c$ and $\chi_{c0}$ case, the QCD contribution also changes the angular
distribution slightly. For example, at $\sqrt{s}=4.26\mathrm{GeV}$, the QED contribution to $\frac{d\sigma}
{d \cos(\theta)}$ is proportional to  $1+4.1\times10^{-3}\cos^2(\theta)$, while the QED+QCD contribution
is proportional to $1+5.8\times10^{-3}\cos^2(\theta)$.

The $\chi_{c1}(2P)$ state has not been observed yet. In some models, the $X(3872)$ is treated as a
mixture of $\chi_{c1}'$ and $D^0\bar{D}^{*0}$ molecule~\cite{Meng-BtoX3872}. Recently, by studying its prompt production cross section at hadron colliders, it was obtained that the size of the $\chi_{c1}'$ component in $X(3872)$
is about $30\%\sim40\%$ \cite{Butenschoen:2013pxa, Meng:2013gga}. If simply choosing $M_{\chi_{c1}
(2P)}=3.872 \mathrm{GeV}$, we predict that
\begin{equation}\label{CS-chic1-2P}
\sigma(e^{+}e^{-}\to \gamma+\chi_{c1}(2P))=0.43 - 0.26\; (\mathrm{pb})
\quad \mathrm{for}\;\; 4.04<\sqrt{s}<5\; \mathrm{GeV.}
\end{equation}
In general, if $\chi_{c1}(2P)$ mass is above the open flavor threshold $M_{D}+M_{D^{\ast}}=3.872\mathrm{GeV}$,
its predominantly decay mode may be $\chi_{c1}(2P)\to DD^{\ast}$ \cite{Eichten:2005ga}. On the
other hand, if $M(\chi_{c1}(2P))<3.872 \mathrm{GeV}$, similar to the $1P$ state case, it will decay mainly into light hadrons, and its total
width will be about one MeV. In some potential model calculations \cite{Li09:ScreenedPotential,Barnes:2005pb},
its E1 transition decay width $\Gamma(\chi_{c1}(2P)\to \gamma+\psi')$ is about $50\sim80~\mathrm{KeV}$.
Based on our calculation and above
analysis, we infer that there are some chances to find the missing $\chi_{c1}(2P)$ state at BESIII whether its mass above
or below the $DD^{\ast}$ threshold  through the $e^{+}e^{-}\to \gamma+\chi_{cJ}(2P)$ process in the continuum
region of $4.04<\sqrt{s}<5~\mathrm{GeV}$.

Particularly, if $X(3872)$ is a mixed state of $\chi_{c1}'$ and $D^0\bar{D}^{*0}$ components, it can also
be detected through the mode $X(3872)\to J/\psi\pi^+\pi^-$, but the
cross section $\sigma(e^{+}e^{-}\to \gamma+X(3872))$ should be smaller than that in (\ref{CS-chic1-2P}) by a factor
of 3 since the probability of $\chi_{c1}'$ in $X(3872)$ is only about $30\%\sim40\%$ \cite{Butenschoen:2013pxa, Meng:2013gga}.

\subsection{$e^{+}e^{-}\to \gamma+\chi_{c2}(nP)$}

The $\chi_{c2}(2P)$ state was observed in the $\gamma\gamma $ collision by Belle Collaboration~\cite{Uehara:2005qd}. Its
mass is about $3.927~\mathrm{GeV}$~\cite{PDG2012}. The QED contributions to the $\gamma+\chi_{c2}(1P)$ and $\gamma+\chi_{c2}(2P)$
cross sections are in the range from $2.5~\mathrm{pb}$ to $0.48~\mathrm{pb}$ and $3.1~\mathrm{pb}$ to $0.55~\mathrm{pb}$,
respectively, for $4.04<\sqrt{s}<5~\mathrm{GeV}$. The QCD contributions are also negative and are about
$-60\%$ of the QED contributions. Using the modified phase space factor, we get that the QED+QCD
contributions are
\begin{eqnarray}
 \sigma(e^{+}e^{-}\to \gamma+\chi_{c2}(nP))=\Big{\{} \begin{array}{ll}
0.48\sim 0.13~ \mathrm{pb}, \quad n=1\\
0.24\sim 0.14~ \mathrm{pb}, \quad n=2
\end{array}
\end{eqnarray}
The QCD contribution changes the angular distribution slightly as well. For example, at
$\sqrt{s}=4.26\mathrm{GeV}$, $\frac{d\sigma}{d \cos(\theta)}$ changes from
$1-9.2\times10^{-2}\cos^2(\theta)$ to $1-9.9\times10^{-2}\cos^2(\theta)$.

In Ref.~\cite{Sang:2012cp}, the production of $\chi_{c2}(1P/2P)$ through indirect photon collision in
$4-5~\mathrm{GeV}$ was studied. After including the one-loop QCD corrections, the cross sections
were found to be a few fb. It is much smaller than those in the $e^{+}e^{-}\to\gamma+\chi_{c2}(1P,2P)$ process.

As references,  in Table II we list the cross sections with the modified phase space factors for $e^{+}
e^{-}\to \gamma+\eta_{c}(1S/2S)(\chi_{c1,2} (1P/2P))$ at some typical energy points in the region
of $\sqrt{s}=4.0\mbox{-}5.0$~GeV, and also show the angular distributions of
$\chi_{c1}$ and $\chi_{c2}$ production at $\sqrt{s}=4.26~\mathrm{GeV}$.

\begin{figure}
\begin{center}
\begin{tabular}{cc}
\includegraphics[scale=0.80]{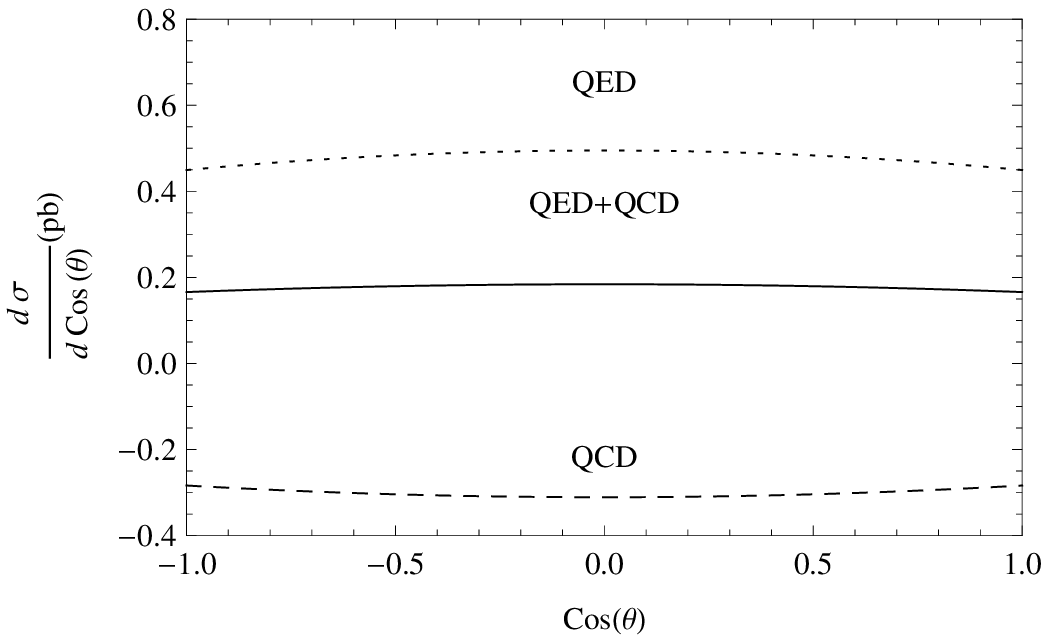}&
\includegraphics[scale=0.80]{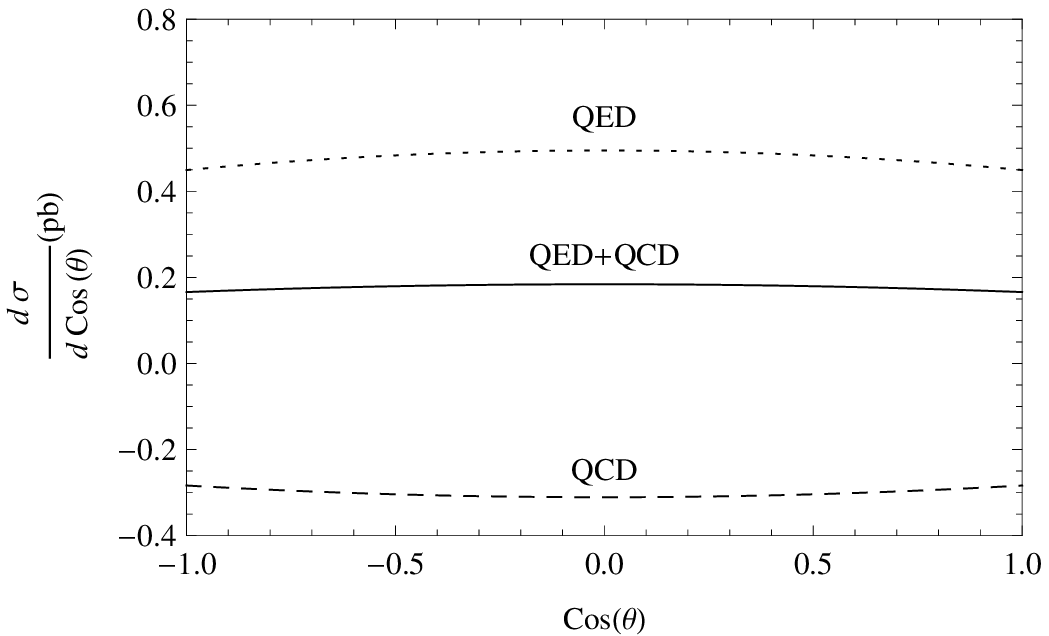}
\end{tabular}
\caption{The angular distributions of $\chi_{c1}$ (left) and $\chi_{c2}$ (right) production in
$e^{+}e^{-}\to \gamma+\chi_{cJ}$ at $\sqrt{s}=4.26 \mathrm{GeV}$. The dotted line denotes the
QED contribution, the dashed line denotes the QCD contribution, and the total QED+QCD contribution
is denoted by the solid line.}
\label{Plot}
\end{center}
\end{figure}

\begin{table}
\begin {center}
\caption{Predicted production cross sections of $\eta_c(1S/2S)$ and $\chi_{c}(1P/2P)$ at
some typical energy points in the region of $\sqrt{s}=4.0\mbox{-}5.0$~GeV.. }
\begin{tabular}{c|ccccccc}
\hline\hline
& \multicolumn{7}{c}{$\sigma/$pb}\\
$\sqrt{s}/GeV$ &~~~$\eta_c$~~~&~~~$\eta^{'}_c$~~~&~~~$\chi_{c0}$~~~&~~~$\chi_{c1}$~~~&~~~
$\chi^{'}_{c1}$~~~&~~~$\chi_{c2}$~~~&~~~$\chi^{'}_{c2}$~~~\\\hline
$4.040$ & $0.91$ & $0.04$ & $~~0.001$ & $0.70$ & $0.32$ & $0.48$ & $0.16$ \\\hline
$4.160$ & $0.86$ & $0.06$ & $-0.005$ & $0.64$ & $0.40$ & $0.41$ & $0.23$ \\\hline
$4.260$ & $0.81$ & $0.08$ & $-0.007$ & $0.58$ & $0.43$ & $0.36$ & $0.24$ \\\hline
$4.360$ & $0.78$ & $0.09$ & $-0.008$ & $0.53$ & $0.43$ & $0.31$ & $0.23$ \\\hline
$4.415$ & $0.76$ & $0.10$ & $-0.008$ & $0.50$ & $0.43$ & $0.28$ & $0.23$ \\\hline
$4.660$ & $0.67$ & $0.13$ & $-0.006$ & $0.40$ & $0.39$ & $0.20$ & $0.19$ \\\hline
$5.000$ & $0.55$ & $0.14$ & $-0.002$ & $0.25$ & $0.26$ & $0.13$ & $0.14$ \\\hline\hline
\end{tabular}
\label{table2}
\end {center}
\vspace{-0.5cm}
\end{table}

\section{Summary}

In summary, we reevaluate the cross sections for $e^{+}e^{-}\to \gamma+\eta_{c}(1S/2S)(\chi_{c1,2} (1P/2P))$
processes at NLO in $\alpha_s$ within the framework of NRQCD factorization at the BESIII energy
region of $\sqrt{s}=4.04\mbox{-}5.0$~GeV. The factorization is verified at this order and the near
threshold effects are partly recovered by using the modified phase space factors for the charmonium states.
The cross sections are as large as $0.1\mbox{-}0.9$ pb, which could be used to search for the missing $2P$
charmonium states or to estimate the continuum backgrounds in the resonance region.

\section*{\large{Acknowledgments}}

This work is supported in part by the National Natural Science Foundation of China
(No 11075002, No 11021092, No 10905001) and the Ministry of Education of China (20100001120013, RFDP).
The work of Zhi-Guo He is supported in part by the German Federal Ministry for Education
and Research BMBF through Grant No.\ 05H12GUE.

$Note\ added$. When this paper was being prepared, a similar calculation was made by Li {\it et al.}~\cite{Zhang2013:e+e-}.
Differing from ours, they use $m_c=M_X/2$ to evaluate the short-distance amplitudes. Thus, the threshold singularities in the
amplitudes, which have been analyzed in Sec.~II, enhance their cross sections for $\chi_c(1P/2P)$ production by
more than a factor of ten near the threshold region.
As for the new BESIII measurement on $\gamma X(3872)$ at $\sqrt{s}=4.229/4.260$
GeV~\cite{BES2013:X3872}, according to our calculation, the cross section of $e^{+}e^{-}\to \gamma X(3872)$ is only 0.15
pb, if the production of $X(3872)$ proceeds dominantly through its $\chi_{c1}(2P)$ component, of which the probability is 0.3-0.4 (see Table II and the context in the subsection III.C). The calculated cross section multiplied by the branching ratio $\mathrm{Br}(X(3872)\to
J/\psi\pi^+\pi^-)$, which is estimated to be about $5\%$~\cite{Butenschoen:2013pxa, Meng:2013gga}, is much smaller than the experimental data at $\sqrt{s}=4.229/4.260$ GeV~\cite{BES2013:X3872}. This suggests that the observed cross section may mainly come from the resonance
contributions through E1 transitions between chamonia~\cite{Meng2013:4260-3872} or some other exotic mechanisms~\cite{Guo2013:4260-3872}.


\end{document}